\documentclass{article}

\usepackage[final,nonatbib]{neurips_2021}

\usepackage[utf8]{inputenc} % allow utf-8 input
\usepackage[T1]{fontenc}    % use 8-bit T1 fonts
\usepackage{hyperref}       % hyperlinks
\usepackage{url}            % simple URL typesetting
\usepackage{booktabs}       % professional-quality tables
\usepackage{amsfonts}       % blackboard math symbols
\usepackage{nicefrac}       % compact symbols for 1/2, etc.
\usepackage{microtype}      % microtypography
\usepackage{xcolor}         % colors
\usepackage{enumitem}
\usepackage{graphicx}

\newcommand{\separateshort}{\vspace{3pt}}
\newcommand{\sys}{\textsf{IRIDEF}}

\newcommand{\discover}{\textsf{DISCOVER}}
\newcommand{\capture}{\textsf{CAPTURE}}
\newcommand{\highlight}{\textsf{HIGHLIGHT}}

\usepackage[linesnumbered,ruled]{algorithm2e}

\title{Interactive Region-of-Interest Discovery using Exploratory Feedback}

\author{Behrooz Omidvar-Tehrani \\ Grenoble AI Institute}

\begin{document}

\maketitle

\begin{abstract}
    In this paper, we propose a geospatial data management framework called \sys\ which captures and analyzes user’s exploratory feedback for an enriched guidance mechanism in the context of interactive analysis. We discuss that exploratory feedback can be a proxy for decision-making feedback when the latter is scarce or unavailable. \sys\ identifies regions of interest (ROIs) via exploratory feedback, and highlights a few interesting and out-of-sight POIs in each ROI. These highlights enable the user to shape up his/her future interactions with the system. We detail the components of our proposed framework in the form of a data analysis pipeline, and present the aspects of efficiency and effectiveness for each component. We also discuss evaluation plans and future directions for \sys.
\end{abstract}

\section{Introduction}

\noindent \textbf{Background.} Nowadays, geospatial data are ubiquitous in various fields of science, such as transportation, smart city management~\cite{RoddickEHPS04,xu2016crowdsourcing}, travel planning~\cite{amer2020interactive}, bike sharing~\cite{chung2018bike}, localized advertising~\cite{feng2016towards}, and regional health-care~\cite{coviz}. A recent solution for an improved geospatial data management is to provide means for {\em interactive analysis}, where users in the loop are guided towards interesting subsets of data in an exploratory iterative manner~\cite{DBLP:conf/cidr/ElMS20,DBLP:journals/pvldb/NandiJ11}. Typically, the guidance is performed through learning user's preferences using a decision-making feedback received from the user in each iteration, e.g., picking (clicking on) a favorite point of interest (POI). However, it is often the case in geospatial scenarios that users forget or don't feel necessary to explicitly express their feedback in what they find interesting. As a result, the interactive dialog will be broken and no guidance can be delivered. In this paper, we focus on the following question: {\em Is it possible to perform interactive analysis on geospatial data without having access to decision-making interactions?}

\separateshort
\noindent \textbf{Proposal.} In the absence of decision-making interactions, we propose to focus on {\em exploratory feedback}, i.e., patterns in signals captured from the user in the background which provide hints on user's interests. For instance, users often hover their mouse (or make frequent touch actions on a touch screen, such as scroll, pinch and zoom) over a region of interest to collect information on the map (e.g., touristic places and hidden gems presented in the form of map layers and tooltips) before landing on a decision about picking a POI in that region, such as a home-stay. Hence it is possible to infer the interest towards that region even without decision-making interactions. This inferred knowledge should be leveraged in the guidance mechanism. An instance of such guidance is to highlight a few interesting POIs in the region of interest. We advocate a geospatial data management framework (called \sys) which captures and analyzes user's exploratory feedback for an enriched guidance mechanism in the context of interactive analysis.

\separateshort
\noindent \textbf{Scenario.} Lindsey is a visiting researcher from the US. She wants to rent a home-stay in Paris via the Airbnb website. She likes to discover the city, hence she is open to any type of lodging in any region with an interest to stay in the center of Paris. Her exploration starts with a query which expresses the preliminary set of her interests. The website returns 1500 different home-stays for her query. While scanning the very first items, she shows (an exploratory) interest towards the region of Trocadero by hovering her mouse around the Eiffel tower and checking the amenities within that region. However, she forgets or doesn't feel the necessity to click a POI (i.e., a home-stay) in that region. While typical recommendation and exploration systems do not necessarily focus on this implicit interest in the future iterations, our framework ensures that Lindsey receives home-stay recommendations related to the Trocadero region even if she didn't provide any decision-making feedback.

\separateshort
\noindent \textbf{Challenges.} Analyzing exploratory feedback is challenging. First, it is not clear how this feedback should be interpreted in terms of the user preferences. Exploratory feedback on geospatial data can be enabled via different signals, such as mouse hovering~\cite{DBLP:conf/edbt/Omidvar-Tehrani20}, touch actions~\cite{DBLP:journals/pvldb/JiangMN13}, voice~\cite{DBLP:conf/cui/ViswanathanGG20}, and gaze~\cite{buscher2012attentive}. Translating such enablers into geospatial semantics is challenging. Second, all exploratory signals are not necessarily useful and some may introduce false positives. For instance, a small mouse move on a typical screen would yield more than 14,000 points (assuming 1600 DPI) which may turn out to be just a random futile move. Beyond the first two challenges, guiding users towards interesting POIs is also challenging, as it requires an exhaustive scan over the geospatial data against the evolving user preferences.

\separateshort
\noindent \textbf{Contributions.} We propose a guidance approach for interactive exploration of geospatial data. Our approach identifies regions of interest (ROIs) without the need for any decision-making feedback. Our proposed guidance mechanism is to highlight a few interesting and out-of-sight POIs in each ROI, and let the user investigate those POIs in his/her future interactions with the system. The following list summarizes the contributions and claims discussed in this paper:

\begin{itemize}[leftmargin=*]
\item We define the notion of ``exploratory user feedback'' which enables a seamless navigation in the geospatial data.
\item We define the notion of ``information highlighting'', a mechanism to highlight important spatial information that is out-of-sight.
% A clear distinction of our proposal with the literature is that the highlighting approach doesn't aim for pruning (such as top-k recommendation), but leveraging the actual data with potentially interesting results.
\item We employ an efficient polygon-based approach to discover ROIs.
\item We propose an approach to compute highlights on-the-fly in an efficient manner.
\end{itemize}

To the best of our knowledge, our contributions have not been investigated before in the  literature. Popular map-based applications such as \textsc{Google Maps} and \textsc{Bing Maps} do not offer interactive functionalities for feedback capturing. In the literature, information highlighting~\cite{Liang2010,Robinson2011,wongsuphasawat2016voyager} and spatial recommendation approaches~\cite{Bao2015,Levandoski:2012} often assume that the user's preferences are static and will never change in time. This limits their functionality for serving the scenarios of an interactive analysis. The process of feedback capturing is mostly formulated for decision-making interactions~\cite{bhuiyan2012interactive,xin2006discovering,dimitriadou2016aide,kamat2014distributed,Omidvar-Tehrani:2015,boley2013one}. While a few fuse decision-making and exploratory feedbacks~\cite{AoidhBW07,Ballatore2008,Liu:2010}, our approach is not dependent on decision-making feedback and is able to operate purely on exploratory feedback. It is to state the obvious that a straightforward extension of our system is to incorporate decision-making feedback (if available) to improve the effectiveness of the system.

\separateshort
\noindent \textbf{Paper outline.} The rest of this paper is organized as follows. In Section~\ref{sec:feedbacks}, we elaborate on different instances of decision-making and exploratory feedbacks in the literature. We discuss the data model and introduce in the problem in Section~\ref{sec:datamodel}. We present our proposed approach in Section~\ref{sec:appr}, and discuss evaluation plans in Section~\ref{sec:eval}. Last, we conclude and present future directions in Section~\ref{sec:conc}.

\begin{figure*}
  \centering
  \includegraphics[width=\columnwidth]{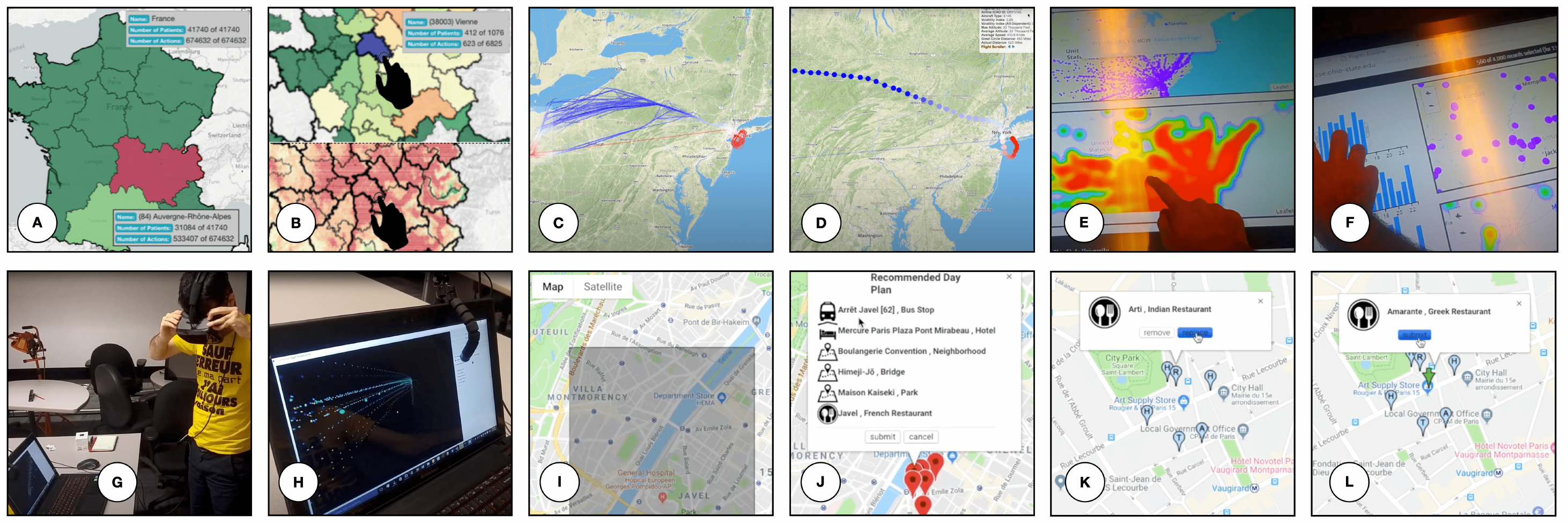}
  \caption{Examples of decision-making and exploratory feedbacks in realistic geospatial scenarios \cite{coviz,DBLP:conf/icde/TehraniNMFY17,amer2020interactive}}
  \label{fig:examples}
\end{figure*}

\section{Decision-making Feedback versus and Exploratory Feedback}
\label{sec:feedbacks}
We briefly discuss a few examples in the literature to clarify the distinction between decision-making and exploratory feedback types in realistic geospatial applications. These examples are illustrated in Figure~\ref{fig:examples}. In summary, we argue that different types of decision-making feedback have been already employed, but the exploratory feedback is often missing.

\separateshort
\noindent \textbf{Medical domain.} \textsc{COVIZ}~\cite{coviz} is an interactive web-based application which enables medical experts to form and compare medical cohorts. In Figure~\ref{fig:examples}-A, the expert clicks on the Auvergne-Rh\^one-Alpes region (as a decision-making feedback) to compare the patient cohort in this particular region with the whole France. In Figure~\ref{fig:examples}-B, the expert adds the air pollution layer to the analysis to examine any potential correlation between the patients' health status and the abundance of the air pollution. While the expert explores the cohort comparisons and pollution correlations, the tool does not collect any exploratory feedback, such as mouse hover and gaze. 

\separateshort
\noindent \textbf{Aviation domain.} \textsc{DV8}~\cite{DBLP:conf/icde/TehraniNMFY17} is an interactive aviation data analysis tool. When several flight trajectories are visualized (Figure~\ref{fig:examples}-C), the expert can click on one trajectory to retrieve its information (departure, destination, etc.), and double-click to solely focus on that single trajectory and analyze it further (Figure~\ref{fig:examples}-D). The interaction is always through the decision-making feedback (single-click and double-click) and the exploratory feedback is not supported. \textsc{DV8} also supports touch gestures, such as pinch and zoom (Figure~\ref{fig:examples}-E) and brush (Figure~\ref{fig:examples}-F). However the touch actions are all considered as decision-making feedback with an immediate resulting action. Hence there is no support for exploratory feedback. The virtual reality (VR) version of \textsc{DV8} (Figures~\ref{fig:examples}-G and~\ref{fig:examples}-H) enhances the exploration experience of the aviation expert, particularly for analyzing flights in different altitudes. While the gaze signal is an exploratory feedback which can be captured through VR, \textsc{DV8} employs the signal only for navigating the geospatial data, and not for guidance.

\separateshort
\noindent \textbf{Travel domain.} \textsc{Simurgh}~\cite{amer2020interactive} is an interactive travel package generation tool. The user can ask for a new day plan using a drag-and-drop action over a region of interest (the drag-and-drop in Figure~\ref{fig:examples}-I and the resulting day plan in Figure~\ref{fig:examples}-J). She can also replace a point of interest by clicking on the point (the selection in Figure~\ref{fig:examples}-K and the replacement in Figure~\ref{fig:examples}-L). All the interactions are defined as the decision-making feedback. In other words, \textsc{Simurgh} does not detect the regions of interest by following the exploratory feedback. 

\section{Data Model and Problem Definition}
\label{sec:datamodel}
To enable feedback capturing, we consider two different layers on a geographical map: {\em spatial layer} and {\em interaction layer}. The spatial layer contains POIs from a spatial database $\mathcal{P}$. The interaction layer contains exploratory feedback points $\mathcal{M}$. These layers are explained below.

\separateshort
\noindent {\bf Spatial layer.} Each POI $p = \langle \mathit{lat}, \mathit{lon} \rangle \in \mathcal{P}$ is described using its geographical coordinates. POIs are also associated to a set of domain-specific attributes $\mathcal{A}$. For instance, in the dataset of a real estate agency, POIs are properties (houses and apartments) and $\mathcal{A}$ contains attributes such as surface, number of rooms and  price. The set of all possible values for an attribute $a \in \mathcal{A}$ is denoted as $dom(a)$. We also define user's feedback $F$ as a vector over all attribute values (i.e., facets), i.e., $F \in \prod_{a \in \mathcal{A}} dom(a)$. The vector~$F$ is initialized by zeros and will be updated to express the user's preferences. The facet-based schema of $F$ ensures that learned feedback is always transparent and interpretable by the user using the facets, and hence reduces algorithmic anxiety~\cite{jhaver2018algorithmic}. 

\separateshort
\noindent {\bf Interaction layer.} We assume that an exploratory signal addresses one specific point $m$ on the screen, e.g., hovering at, gazing at, or providing a voice command about $m$. When an exploratory signal is received, the point $m$ is appended to the set $\mathcal{M}$. Each point is a tuple $m = \langle x, y, t \rangle$, where $x$ and $y$ specify the affected pixel location and $t$ is a timestamp. To conform with geographical standards, we assume $m = \langle 0, 0, t\rangle$ sits at the middle of the interaction layer, both horizontally and vertically, for any $t$.
% We also define a set of regions $\mathit{R}$ where each $r \in \mathit{R}$ contains a set of mouse move points $\mathcal{M}_r \subseteq \mathcal{M}$ and a time interval $T_r$ where $\forall m=\langle x,y,t\rangle \in \mathcal{M}_r, t \in T_r$.

\separateshort
\noindent {\bf Transitioning between the layers.} The user is in contact with the interaction layer. To update the feedback vector $F$, we need to translate pixel locations in the interaction layer to latitudes and longitudes in the spatial layer. 
% While there is no precise transformation from planar to spherical coordinates,
We employ equirectangular projection to obtain the best possible approximation of a point $m = \langle x,y,t \rangle \in \mathcal{M}$ in the spatial layer, denoted as $p(m)$. 
% describes this formula to transform a point $m = \langle x,y,t \rangle$ in the interaction layer to a spatial point $p = \langle lat, lon \rangle$ in the spatial layer. Note that the resulting $p$ is not necessarily a member of $\mathcal{P}$. 

\begin{equation}\label{eq:equirectangular}
p(m = \langle x,y,t \rangle) = \langle \mathit{lat}=y + \gamma, \mathit{lon}=\frac{x}{\mathit{cos}\gamma} + \theta\rangle
\end{equation}

The inverse operation, i.e., transforming a point $p = \langle \mathit{lat}, \mathit{lon}\rangle$ from the spatial layer to the interaction is done using Equation~\ref{eq:reverse}.

\begin{equation}\label{eq:reverse}
m(p = \langle \mathit{lat}, \mathit{lon}\rangle) = \langle x= (\mathit{lon} - \theta) \times \mathit{cos}\gamma,  y = \mathit{lat} - \gamma\rangle
\end{equation}

% \separateshort
The reference point for the transformation is the center of both layers. In Equations~\ref{eq:equirectangular} and~\ref{eq:reverse}, we assume that $\gamma$ is the latitude and~$\theta$ is the longitude of a point in the spatial layer corresponding to the center of the interaction layer, i.e., $m= \langle 0,0 \rangle$.

\separateshort
\noindent {\bf Problem definition.} Given the user's feedback $F$, we are interested in solving two consecutive problems: $(i)$ discover regions of interest in the form of geospatial clusters whose centroids correlate with $F$ (with respect to the POI attributes in which the user is interested in), and $(ii)$ for each discovered region, find at most $k$ POIs ($k$ is an input parameter) which are relevant to~$F$ and have high exploration quality. We define relevance and exploration quality in Section~\ref{sec:appr}.

\begin{figure}[h]
\centering
\includegraphics[width=0.75\columnwidth]{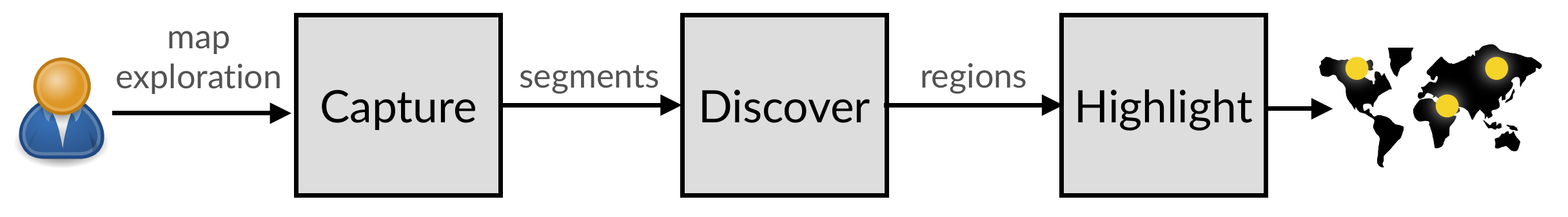}
\caption{\sys\ framework.}
 \label{fig:sys}
\end{figure}

\section{Proposed Approach}
\label{sec:appr}
We propose \sys\ (\textbf{I}nteractive \textbf{R}egion-of-\textbf{I}nterest \textbf{D}iscovery using \textbf{E}xploratory \textbf{F}eedback), a framework for exploiting exploratory feedback to highlight interesting POIs as future analysis directions. As depicted in Figure~\ref{fig:sys}, our approach consists of a pipeline with three main components: \capture, \discover, and \highlight. After the user has explored the map for a while, \sys\ captures exploratory feedback from the exploration (i.e., the \capture\ component detailed in Section
~\ref{sec:capture}). Then a set of regions of interest (ROIs) will be discovered using the captured feedback (i.e., the \discover\ component detailed in Section~\ref{sec:discover}). Finally some out-of-sight interesting POIs will be highlighted for each discovered ROI (i.e., \highlight component detailed in Section~\ref{sec:highlight}). In the following, we first discuss the desiderata behind our approach, and then detail each component of the pipeline.

\subsection{Principles}
\label{sec:principles}
In order to maximize the usability of \sys, we believe that the framework should be generic and fluid, as discussed below.

\separateshort
\noindent \textbf{Genericness.} \sys's pipeline is applicable to different datasets and different types of exploratory feedback. This enables \sys\ to cover different exploration scenarios. The minimal requirement is that the input dataset and the feedback signal match with our data model (Section~\ref{sec:datamodel}).

\separateshort
\noindent \textbf{Fluidity.} A fluid interactive system does not break the user's train of thought. The fluidity is ensured by rendering results in an {\em efficient} and {\em effective} manner. In the \capture\ component, effectiveness is satisfied by disregarding irrelevant signals. In the \discover\ and \highlight\ components, effectiveness is interpreted as delivering meaningful and useful regions (ROIs) and highlights (POIs), respectively. In all of the components, efficiency is to return results instantaneously, often considered to be $\leq 500ms$~\cite{fekete2016progressive}.

\subsection{CAPTURE Component}
\label{sec:capture}
Exploratory feedback can be captured using different latent signals, e.g., time dedicated to item details, touch actions, gaze, mouse moves, scrolling speed, etc. Without loss of generality, we focus on mouse moves as an instance of exploratory feedback signal. A particular challenge in capturing mouse moves as the exploratory feedback is that the user may mindlessly move the mouse everywhere on the map. Obviously, this should not signify that all the locations are equally important to the user. An {\em effective} approach should only capture a subset of this feedback which is then useful for discovering ROIs. Also an {\em efficient} approach should capture this feedback without any interruption in the fluidity of the user experience. For an effective and efficient feedback capturing, \sys\ performs the two following actions:

\begin{enumerate}[leftmargin=*]
\item First, it records the exploratory signals (by adding the coordinates of the screen points they were applied on to $\mathcal{M}$) only every $\varepsilon$ milliseconds to prevent adding redundant points.
\item After a given period of feedback capturing time, it groups the recorded signals into $g$ different segments, $\mathcal{M}_1$ to $\mathcal{M}_g$. The first segment starts at time zero (where the system started to operate), and the last segment ends at the current time. 
\end{enumerate}

% \separateshort
% \noindent {\bf Feedback capturing parameters.} 
The choice of $\varepsilon$ depends on various parameters such as the application (e.g., tourism, delivery, transportation) and the user's expertise. For instance, a larger~$\varepsilon$ seems more appropriate for novice users, as they might perform many random moves to get acquainted with the data. In conformance with progressive data analytics~\cite{fekete2016progressive}, we set $\varepsilon = 100ms$ as the default value to ensure continuity preserving latency.

\begin{algorithm}[t]
\DontPrintSemicolon
\KwIn{Mouse move points $\mathcal{M}$, time gap $\varepsilon$, segmentation strategy $\psi$}
\KwOut{Segments $\mathcal{M}_i$, $i \in [1,g]$}
$\mathit{segment\_count} \gets 0$\;
\For{$m \in \mathcal{M}$ captured every $\varepsilon$ milliseconds}
{
$\mathcal{M}[\mathit{segment\_count}] \gets \mathcal{M}[\mathit{segment\_count}] \cup \{m\}$\;
$\mathit{segment\_change} \gets \mathit{check\_strategy}(\psi,m,\mathcal{M})$\;
\If{$\mathit{segment\_change} = \mathit{true}$}
{
$\mathit{segment\_count} \gets \mathit{segment\_count} + 1$\;
$\mathcal{M}[\mathit{segment\_count}] \gets \emptyset$\;
}
}
\Return{$\mathcal{M}_i$, $i \in [1,g]$ where $g = \mathit{segment\_count}$}\; 
\caption{\capture\ algorithm}
\label{algo:capture}
\end{algorithm}

\separateshort
% \noindent {\bf Segmentation.} 
Moreover, the end of a segment is determined by one of the following approaches:

\begin{itemize}[leftmargin=*]
\item $\psi_1$: End the current segment after a fixed amount of time (i.e., fixed-length segments). In this case, the value of $g$ is selected based on the spatial density of the dataset under investigation.
\item $\psi_2$: End the current segment if the mouse location is unchanged for a certain amount of time.
\item $\psi_3$: End the current segment after a drastic change in the signal, where the drift is captured using signal segmentation approaches. We employ the Wedding Cake technique for the dynamic segmentation of our signals~\cite{krumm2006predestination,moosavi2017characterizing}.
\end{itemize}

% \separateshort
Algorithm~\ref{algo:capture} summarizes the \capture\ process.

\subsection{DISCOVER Component}
\label{sec:discover}
The objective of this step in the \sys\ pipeline is to obtain one or several ROIs in which the user has expressed his/her exploratory feedback. We conjecture that a region is more interesting for the user if it is denser, i.e., the user moves the mouse in that region frequently, to collect information from the background map. Hence ROIs can be simply discovered as dense clusters of mouse move points. We denote the set of all ROIs as $\mathcal{R}$ and we refer to the $i$-th ROI as $R_i \in \mathcal{R}$. Algorithm~\ref{algo:discover} summarizes the \discover\ process.

\separateshort
% \noindent {\bf Clustering segments.} 
We employ ST-DBSCAN~\cite{Birant:2007}, a space-aware variant of DB-SCAN, to cluster points in each segment (line~\ref{ln:dbscan} in Algorithm~\ref{algo:discover}). For each subset of mouse move points $\mathcal{M}_i$, $i \in [1,g]$, ST-DBSCAN begins with a random point $m_0 \in \mathcal{M}_i$ and collects all density-reachable points from $m_0$ using a distance metric. As mouse move points are in the 2-dimensional pixel space (i.e., the screen), we choose euclidean distance as the distance metric. A density-reachable point $m_i$ is either directly reachable from $m_0$, i.e., the distance between $m_i$ and $m_0$ is lower than a distance threshold (an input parameter for the ST-DBSCAN algorithm), or reachable via a path $m_0 \dots m_{j-1}, m_j \dots m_i$ where each point $m_j$ in the path is directly reachable from its immediately prior point in the path $m_{j-1}$. If $m_0$ turns out to be a core point, a cluster will be generated. A point is core if there exist a certain amount of points in its vicinity, i.e., with a distance lower than the distance threshold. The minimum number of points for a core point is yet another input parameter for ST-DBSCAN. If $m_0$ is not a core point, the algorithm picks another random point in $\mathcal{M}_i$. The process is repeated until all points have been processed. We denote the set of all resulting clusters for $\mathcal{M}_i$ as $\mathcal{C}_i = \{C_1, C_2, \dots\}$.

\begin{algorithm}[t]
\DontPrintSemicolon
\KwIn{Segments $\mathcal{M}_1$ to $\mathcal{M}_g$, user feedback vector $F$, number of interactions performed so far $T$}
\KwOut{Set of discovered ROIs $\mathcal{R}$}
$O \gets \emptyset$\tcp*[l]{the set of all polygons initialized as empty}
$\mathcal{R} \gets \emptyset$\tcp*[l]{the set of all ROIs initialized as empty}
\For{each segment $\mathcal{M}_i$}
{
$\mathcal{C}_i \gets \mathit{ST\_DBSCAN}(\mathcal{M}_i)$\label{ln:dbscan} \tcp*[l]{all clusters inside $\mathcal{M}_i$}
$\mathcal{C}_i \gets \mathit{AklToussaint}(\mathcal{C}_i)$\label{ln:toussaint} \;
$O_i \gets \mathit{Graham\_scan}(\mathcal{C}_i)$\label{ln:graham} \tcp*[l]{all polygons inside $\mathcal{M}_i$}
$O_i.\mathit{expand}(\mathit{confidence}(F,T))$\label{ln:expand} \tcp*[l]{Equation~\ref{eq:confidence}}
$O \gets O \cup O_i$
}
\For{each pair of polygons $O_x \in O$ and $O_y \in O$\label{ln:startroi}}
{
$S \gets \mathit{intersect}(O_x,O_y)$\;
\lIf{$S.\mathit{size} > 0$}{$\mathcal{R} \gets \mathcal{R} \cup \{S\}$}
}\label{ln:endroi}
\Return{$\mathcal{R}$}\; 
\caption{\discover\ algorithm}
\label{algo:discover}
\end{algorithm}

\separateshort
% \noindent {\bf Discovering ROIs.} 
Once the clusters are obtained for all the subsets of~$\mathcal{M}$, we find their intersections to locate recurring regions. Note that we don't aim to directly consider the clusters $\mathcal{C}$ as the ROIs, as they may contain noisy signals. Their intersection counts as a confirmation of user preferences. To obtain intersections, we need to clearly define the spatial boundaries of each cluster. For this aim, we discover the polygons which cover the points inside each cluster. We employ Graham scan algorithm (line~\ref{ln:graham} in Algorithm~\ref{algo:discover}) which is an efficient method to compute the convex hull for a given set of points in a 2D plane~\cite{graham1972efficient}. We reduce the typical complexity of Graham scan (i.e., $\mathcal{O}(|C_i| \times \mathit{log} |C_i|)$, $|C_i|$ being the number of points in the $i$-th cluster) to $\mathcal{O}(|C_i|)$ by ordering the cluster members by their spatial coordinates. For more efficiency, we perform Akl-Toussaint heuristics~\cite{devroye1981note} before the polygon computation to prune the points which are unnecessary for shaping the polygons (line~\ref{ln:toussaint} in Algorithm~\ref{algo:discover}). The intersections between the polygons constitute the ROIs (lines~\ref{ln:startroi} to~\ref{ln:endroi} in Algorithm~\ref{algo:discover}).

\separateshort
\noindent {\bf Personalizing discovered ROIs.} By default, our ROI discovery approach creates strictly tight ROIs, i.e., the area of the polygons is exactly inferred by the points it covers. However in exploratory scenarios, the feedback points do not necessarily reflect the exact interests of the user. The user exposes his/her interests in a gradual manner using exploratory feedback captured in several iterations. We believe that the user's confidence (interpreted as the richness of the user feedback vector $F$) should impact the way ROIs are computed, hence personalized ROIs. In case the user is less confident (e.g., the user is in early stages of his/her exploration), ROIs should be expanded in their area (up to twice their original size) to let more opportunities arise (line~\ref{ln:expand} in Algorithm~\ref{algo:discover}). The user confidence is computed as follows.

\begin{equation}\label{eq:confidence}
    \mathit{confidence}(F,T) = \mathit{min}(1.0,\frac{||F||_0}{\xi \times T})
\end{equation}

In Equation~\ref{eq:confidence}, $\xi$ is a feedback frequency, and $T$ is the number of interactions performed so far. For instance, given $|F| = 50$, $T = 10$, and assuming that a typical user provides $7$ exploratory signals per iteration, the confidence will be equal to $0.71$. The confidence is a coefficient for stretching the ROI area. Let $A_1$ denote the area of the ROI $R_1$, the confidence-aware area $A'_1$ is computed as follows: $A'_1 = (A_1 + A_1 \times \mathit{confidence})$. This process is shown in line~\ref{ln:expand} of Algorithm~\ref{algo:discover}.

\begin{figure*}[t]
\includegraphics[width=\textwidth]{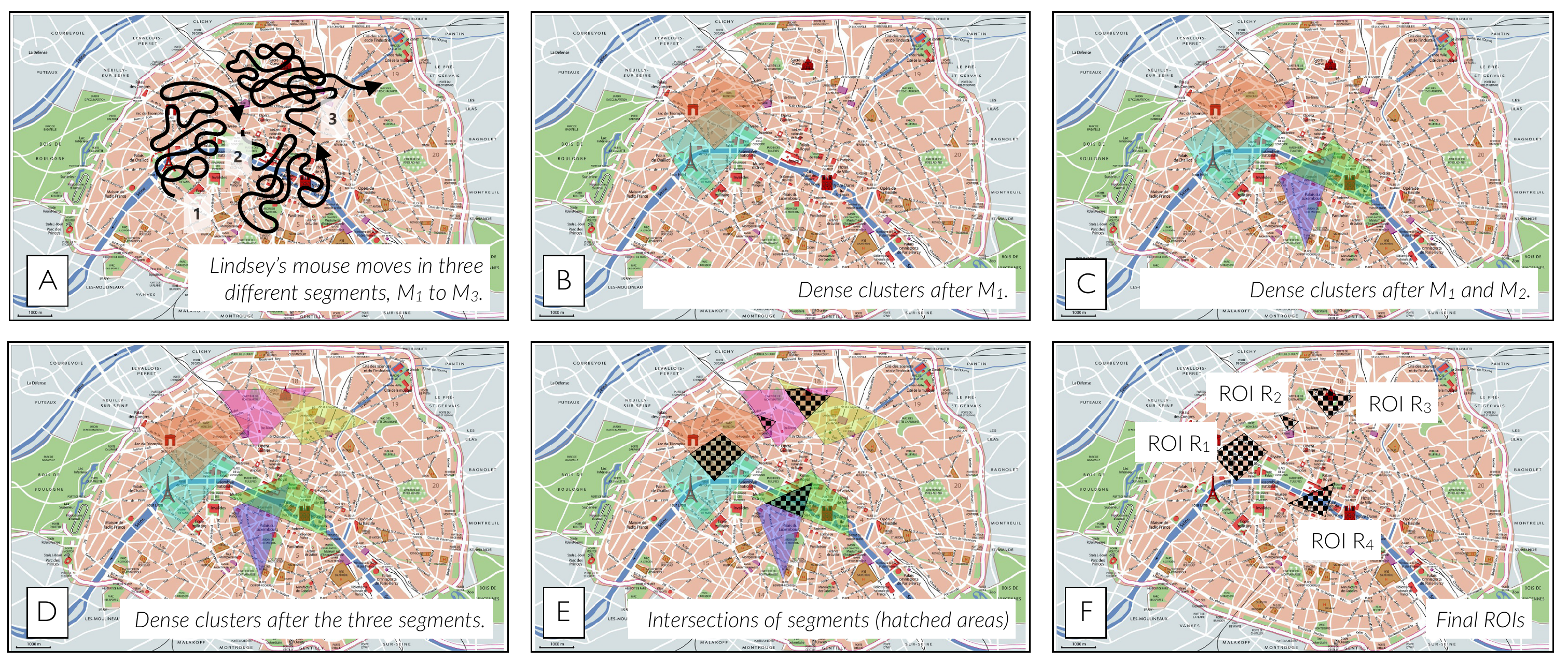}
\caption{An example of discovering ROIs~\cite{DBLP:conf/edbt/Omidvar-Tehrani20}.}
\label{fig:example}
\end{figure*}

\separateshort
\noindent {\bf Example.} Figure~\ref{fig:example} shows the steps that Lindsey follows to explore home-stays in Paris. For the sake of simplicity, we assume Lindsey's confidence is $1.0$. Figure \ref{fig:example}.A shows the mouse moves of Lindsey in different time stages. In this example, we consider $g = 3$ and capture Lindsey's feedback in three different time segments  with fixed-length, i.e., $\psi_1$ (progressing from Figures \ref{fig:example}.B to \ref{fig:example}.D). It shows that Lindsey started her search around Eiffel Tower and Arc de Triomphe (Figure \ref{fig:example}.B) and gradually showed interest in areas located south (Figure \ref{fig:example}.C) and north (Figure \ref{fig:example}.D) as well. All intersections between those clusters are discovered~(hatched regions in Figure \ref{fig:example}.E) which will contribute to the set of interesting regions (Figure \ref{fig:example}.F), i.e., ROI1 to ROI4.

\subsection{HIGHLIGHT Component}
\label{sec:highlight}
We define highlights as a subset of POIs in the form of suggestions for directions of future analysis of the user. The highlights are generated by performing the three following steps: {\em matching points}, {\em updating feedback}, and {\em highlighting POIs}. First, we find POIs which fit into the polygons obtained in the \discover\ component. Then we update the user feedback $F$ according to those POIs. Finally we highlight a set of POIs based on the updated content of $F$.

% We also mention peculiarity as a control mechanism for $k$.
\separateshort
\noindent \textbf{Matching points.} Being a function of mouse move points, ROIs are discovered in the interaction layer. We then need to find out which POIs in $\mathcal{P}$ fall into ROIs. We employ Equation~\ref{eq:reverse} to transform those POIs from the spatial layer to the interaction layer. Then a simple spatial containment function can verify whether a given POI fits into a given ROI.\footnote{\it Typically, we use the implementation of $\mathit{ST\_Within}()$ module in PostGIS for the containment verification.} To improve efficiency, we employ Quadtrees~\cite{finkel1974quad} in a two-step approach: $(i)$ In an offline process, we build a Quadtree index for all POIs in~$\mathcal{P}$. We record the membership relations between POIs and Quadtree grid cells in the index. $(ii)$ Once ROIs are discovered, we record which cells in the Quadtree index intersect with the ROIs.  For matching POIs, we only check a subset which is inside the cells associated to ROIs and ignore the ones outside, hence a drastic pruning of POIs in~$\mathcal{P}$. Given an ROI $R_i$, we denote the set of its matching points as~$\mathcal{P}_{i}$. We also define the binary vector $\overrightarrow{\mathcal{P}_{i}}$ whose cell of $\langle a_j, v_w \rangle$ is $1$ if at least one point in $\mathcal{P}_{i}$ gets the value $v_w \in \mathit{dom}(a_j)$ for the attribute $a_j \in \mathcal{A}$, otherwise $0$.

\separateshort
\noindent \textbf{Updating feedback.} The matching points depict the exploratory preferences of the user. To memorize these preferences, we update the feedback vector $F$ using the attributes of the matching points. We consider an increment value $\delta$ to update $F$. If $p$ is a matching point and gets $v_w \in \mathit{dom}(a_j)$ for attribute $a_j \in \mathcal{A}$, we augment the value in the $F$'s cell of $\langle a_j, v_w \rangle$ by the factor $\delta$. Note that we only consider incremental feedback, i.e., we never decrease a value in~$F$. The vector $F$ will become normalized after each update using a softmax function. The updated feedback vector is fully transparent and the user can easily apprehend what has been learned from his/her previous actions. Our current update model considers the feedback vector to be recency-agnostic. We leave the integration of recency as future work.

\separateshort
\noindent \textbf{Highlighting POIs.} The updated feedback vector $F$ is the input to the highlighting phase. The objective is to select $k$ POIs out of all POIs inside ROIs whose relevance and exploration quality are maximal. We denote the set of highlights as $\mathcal{H}$. We propose two approaches to achieve our objective, depending on how we define relevance and quality:

\begin{algorithm}[t]
\DontPrintSemicolon
\KwIn{Discovered ROIs $\mathcal{R}$, user feedback vector $F$, $k$, $\mathit{time\_limit}$, $\mathit{similarity\_threshold}$}
\KwOut{Highlights $\mathcal{H}$}
$\mathcal{H} \gets \emptyset$\tcp*[l]{highlights}
\For{each discovered ROI $R_i \in \mathcal{R}$}
{
$\mathcal{P}_i \gets \mathit{match\_points}(R_i)$\;
$F.\mathit{update}(\mathcal{P}_i)$\;
$\mathcal{L}_i \gets $ sort the POIs in $\mathcal{P}_i$ in decreasing order of their similarity with $F$\label{ln:inverted}\;
$p^* \gets \mathit{most\_similar\_point}(\mathcal{P}_i,F)$\;
$k' \gets k \times \mathit{peculiarity}(R_i)$\label{ln:pec1}\;
$\mathcal{H}[R_i] \gets \mathit{top}(\mathcal{L}_i, k')$\label{ln:init}\;
$p_{\mathit{next}} \gets \mathit{get\_next}(\mathcal{L}_i)$\;
\While{$\mathit{time\_limit}$ not exceeded and $\mathit{similarity}(p_{\mathit{next}}, p^*) \leq \mathit{similarity\_threshold}$}
{
\For{$p_{\mathit{current}} \in \mathcal{H}[R_i]$}
{
\lIf{$\mathit{diversity\_improved}(\mathcal{H}[R_i], p_{\mathit{next}}, p_{\mathit{current}})$\label{ln:divimprove}}
{
$\mathcal{H}[R_i] \gets \mathcal{H}[R_i] \cup \{p_{\mathit{next}}\} \setminus \{p_{\mathit{current}}\}$
}
}
$p_{\mathit{next}} \gets \mathit{get\_next}(\mathcal{L}_i)$\;
}
}
\Return{$\mathcal{H}$}
\caption{Greedy \highlight\ algorithm}
\label{algo:greedy}
\end{algorithm}

\separateshort
\noindent \textit{Greedy approach.} Inspired from~\cite{DBLP:conf/edbt/Omidvar-Tehrani20,DBLP:conf/ithings/Omidvar-Tehrani17,behroozsigspatial}, we define the relevance as the Cosine similarity between~$F$ and the POIs (note that the feedback vector $F$ and the POIs are defined over the same schema), and the quality as the diversity between the POIs. The diversity is computed using Cosine distance between the POI attribute values. We then follow a greedy approach for each ROI to maximize diversity while respecting a lower bound on similarity. Algorithm~\ref{algo:greedy} summarizes this approach. The similarity values are preprocessed and organized in $\mathcal{L}_i$ for all POIs in $\mathcal{P}_i$ (line~\ref{ln:inverted} in the algorithm). The algorithm starts the greedy process by initializing a list $\mathcal{H}[R_i]$ with $k'$ POIs at the top of $\mathcal{L}_i$, i.e., the most similar POIs in $\mathcal{P}_i$ to $F$ (line~\ref{ln:init} in the algorithm). While a time limit is not exceeded (time limit is an input parameter which is often set to values $\leq 500ms$~\cite{fekete2016progressive}), the algorithm scans $\mathcal{L}_i$ sequentially to find appropriate POI replacements in $\mathcal{H}[R_i]$ to improve diversity (line~\ref{ln:divimprove} of the algorithm). Once the greedy loop is done, the set $\mathcal{H}$ will be returned by the algorithm, containing the highlights for all the discovered ROIs.

\separateshort
\noindent \textit{Fuzzy approach.} Inspired from~\cite{leroy2015building,DBLP:conf/dsaa/SinghBHAE17,amer2019grouptravel,amer2020interactive}, we employ fuzzy clustering to process all ROIs simultaneously. Algorithm~\ref{algo:fuzzy} summarizes this approach. The relevance is defined in the same way as the greedy approach, and the exploration quality is defined using two factors: cohesiveness between POIs of the same ROI (opposite of diversity, hence measured using Cosine similarity), and representativeness, i.e., the sum of euclidean distances between ROI centroids. We use a weighted sum over relevance and quality where the weights are user-defined parameters ($w_1$ to $w_3$ in line~\ref{ln:weights} of Algorithm~\ref{algo:fuzzy}). Through several trial-and-error tests and user studies in previous works~\cite{amer2019grouptravel,DBLP:conf/dsaa/SinghBHAE17}, we found that the most ideal set of weights are $w_1 = 0.5$, $w_2= 0.25$ and $w_3=0.25$. The algorithm refines the centroids of ROIs iteratively until convergence (lines~\ref{ln:startconv} to~\ref{ln:endconv} in Algorithm~\ref{algo:fuzzy}). Then $k'$ most probable points (in fuzzy clustering semantics) will be returned as highlights for each centroid (line~\ref{ln:done} in Algorithm~\ref{algo:fuzzy}).

\separateshort
\noindent \textit{Which approach to choose?} We conjecture that the greedy approach is more appropriate for the bird's-eye view exploration, which mainly refers to early stages of the exploration where the user is trying to get acquainted with the geospatial data by random explorations. In this case, ROIs do not necessarily need to be related and may represent independent future directions. However, in the case of more focused exploration scenarios, the fuzzy approach would be able to deliver highlights with more coverage over the whole regions of interest. We plan to validate these hypotheses via extensive qualitative evaluations.

\separateshort
\noindent \textbf{Peculiar highlighting.} Recall the main objective of the highlighting component is to return out-of-sight POIs as future analysis directions. This simply means that the neighborhoods that have been already investigated by the user are less peculiar, and the POIs within those regions may not be as interesting as the ones in unexplored regions. Given an ROI $R_i$, we define its peculiarity score as follows.

\begin{equation}\label{eq:pecu}
    \mathit{peculiarity}(R_i) = \mathit{Cosine\_similarity}(F,\overrightarrow{\mathcal{P}_i})
\end{equation}

\begin{algorithm}[t]
\DontPrintSemicolon
\KwIn{Discovered ROIs $\mathcal{R}$, user feedback vector $F$, $k$}
\KwOut{Highlights $\mathcal{H}$}
$\mathcal{P}_{\mathit{all}} \gets \emptyset$\;
\For{each discovered ROI $R_i \in \mathcal{R}$}
{
$\mathcal{P}_{\mathit{all}} \gets \mathcal{P}_{\mathit{all}} \cup \mathit{match\_points}(R_i)$\;
$\mathit{centroid}_{\mathit{old}} \gets \emptyset$\;
$\mathit{centroid}_{\mathit{current}} \gets \mathit{get\_centroid}(R_i)$\;
}
$k' \gets k \times \mathit{peculiarity}(R_i)$\label{ln:pec2}\;
\While{$\delta(\mathit{centroid}_{\mathit{old}}, \mathit{centroid}_{\mathit{current}})$ is significant\label{ln:startconv}}
{
$\mathit{centroid}_{\mathit{old}} \gets \mathit{centroid}_{\mathit{current}}$\;
$\mathit{centroid}_{\mathit{current}} \gets \mathit{argmax}_{k'}(w_1 \times \mathit{relevance}(\mathcal{P}_{\mathit{all}}),$ $w_2 \times \mathit{cohesiveness}(\mathcal{P}_{\mathit{all}}),$ $w_3 \times \mathit{representativeness}(\mathcal{P}_{\mathit{all}}))$\label{ln:weights}\;
}\label{ln:endconv}
$\mathcal{H} \gets \mathit{fuzzy\_clusters}(\mathit{centroid}_{\mathit{current}})$\label{ln:done}\;
\Return{$\mathcal{H}$}
\caption{Fuzzy \highlight\ algorithm}
\label{algo:fuzzy}
\end{algorithm}

\separateshort
We then enrich the traditional $k$ parameter with the peculiarity semantics as follows: $k' = \lfloor k \times \mathit{peculiarity}(R_i) \rfloor$ (line~\ref{ln:pec1} in Algorithm~\ref{algo:greedy} and line~\ref{ln:pec2} in Algorithm~\ref{algo:fuzzy}). Note that $k'$ is the peculiarity-aware version of the $k$. This simply means that $k'$ is lower for less peculiar ROIs, and hence less POIs will be highlighted in them. For instance, in case $F$ has already captured feedback about two-bedroom home-stays and an ROI has only amenities with two bedrooms, that ROI will receive a low peculiarity score, and hence very few POIs will be highlighted in it.

\section{Discussion on Evaluation}
\label{sec:eval}
We plan to perform the following evaluation strategies to validate the usefulness of \sys: 

\separateshort
\noindent \textbf{Single-shot quantitative analysis.} Although our approach is multi-shot, we can consider only one iteration of our approach (\capture\ $\rightarrow$ \discover\ $\rightarrow$ \highlight) and see how the components behave in this single iteration. The behavior can be captured through execution time and memory consumption, as well as precision. We average over several single-shot runs. The feedback will be captured through crowdsourcing campaigns.

\separateshort
\noindent \textbf{Simulation study.} We simulate interactive scenarios using virtual agents and measure accumulated quality such as precision, hit ratio, and diversity.

\separateshort
\noindent \textbf{User study.} We also perform an in-depth lab study and an in-breadth crowdsourcing study to survey real users about their perception on the resulting regions (ROIs) and the highlights (POIs).

\section{Conclusion and Future Work}
\label{sec:conc}
In this paper, we present \sys, an approach to interactively discover regions of interest (ROIs) using exploratory feedback. The exploratory feedback is captured from mouse moves over the geographical map while analyzing spatial data. We propose a novel polygon-based mining algorithm which returns a few highlights (POIs) in conformance with user's exploratory preferences. The highlights enable users to have a better understanding of what to focus on in the followup steps in their analysis scenarios. We plan to extend \sys\ in several directions, such as the incorporation of multi-modal exploratory feedback and the generation of sequential highlights as a mobility-aware guidance.

\section*{Acknowledgment}
The author thanks Thibaut Thonet, Sruthi Viswanathan, Fabien Guillot, Jean-Michel Renders, and Placido Neto for their constructive comments in the process of writing this paper.

\bibliographystyle{unsrt}
\bibliography{ref}

\end{document}